\documentclass[%
 reprint,
 superscriptaddress,
 amsmath,amssymb,
 aps,
 prl,
 floatfix,
 nobibnotes
]{revtex4-2}

\usepackage{graphicx}
\usepackage{dcolumn}
\usepackage{bm}
\usepackage{xcolor}
\usepackage{hyperref}
\hypersetup{
    pdfstartview={FitH},
    colorlinks=true,
    linkcolor=blue,
    citecolor=blue,
    filecolor=blue,
    urlcolor=blue 
}

\usepackage{pdfpages}
\usepackage{pgffor}
\makeatletter
\AtBeginDocument{\let\LS@rot\@undefined}
\makeatother

\begin{document}

\title{Detecting vortex motion through spatially correlated nonequilibrium noise}

\author{Yifan F. Zhang}
\email{yz4281@princeton.edu}
\affiliation{Department of Electrical and Computer Engineering, Princeton University, Princeton, NJ 08544, USA}
\author{Rhine Samajdar}
\email{rhine\_samajdar@princeton.edu}
\affiliation{Department of Electrical and Computer Engineering, Princeton University, Princeton, NJ 08544, USA}
\affiliation{Department of Physics, Princeton University, Princeton, NJ 08544, USA}
\author{Sarang Gopalakrishnan}
\affiliation{Department of Electrical and Computer Engineering, Princeton University, Princeton, NJ 08544, USA}

\begin{abstract}
Resistive transport near a superconducting phase can arise from the motion of normal-state quasiparticles or that of vortices. The conductivity alone does not distinguish between these mechanisms. 
We propose an unambiguous method for telling them apart, using the recently developed experimental tool of covariance magnetometry, which uses nitrogen-vacancy centers in diamond to probe real-time spatiotemporal correlations in magnetic noise.
Our key insight is that, under an applied current, the underlying charge carriers leave a directional fingerprint in the spatially correlated magnetic noise above the sample: ordinary electric carriers drift parallel to the current, whereas vortices, owing to the Magnus force, drift perpendicular to it. 
The noise covariance detects this anisotropy and identifies the vortex-driven nature of transport. 
We compute the noise correlations expected for a representative thin-film superconductor and demonstrate that the anisotropic signal is well within the reach of current experimental capabilities.

\end{abstract}

\maketitle

In the low-temperature limit, normal metals are Fermi liquids~\cite{nozieres2018theory}, in which the dynamical variables are electron-like quasiparticles. Recent experimental work has yielded a diverse collection of \emph{strange} or \emph{anomalous} metals~\cite{KapitulnikKivelsonSpivak2019}, which are also resistive but behave in ways inconsistent with Fermi-liquid theory. Identifying the dynamical variables in strange metals, and how they give rise to transport, remains an open problem---and also one that linear-response measurements alone cannot settle, since they do not directly probe the nature of charge carriers. The development of experimental tools capable of going beyond linear response, and probing quantities like high-order correlation functions or nonequilibrium noise, naturally raises the question of whether they can distinguish between proposed theories of the strange metal (and related regimes like the pseudogap in cuprate superconductors~\cite{sachdev2025foot, sachdev2009quantum}). Understanding what such probes reveal about strange metals is an active effort, both experimentally and theoretically~\cite{Chen2023, zhang2025evidence, wang2026reconciling, Wang2022shot, Nikolaenko2023, WuFoster2024, Green2006, Gopalakrishnan2024non}.

\begin{figure}[b]
    \centering
    \includegraphics[width=\linewidth]{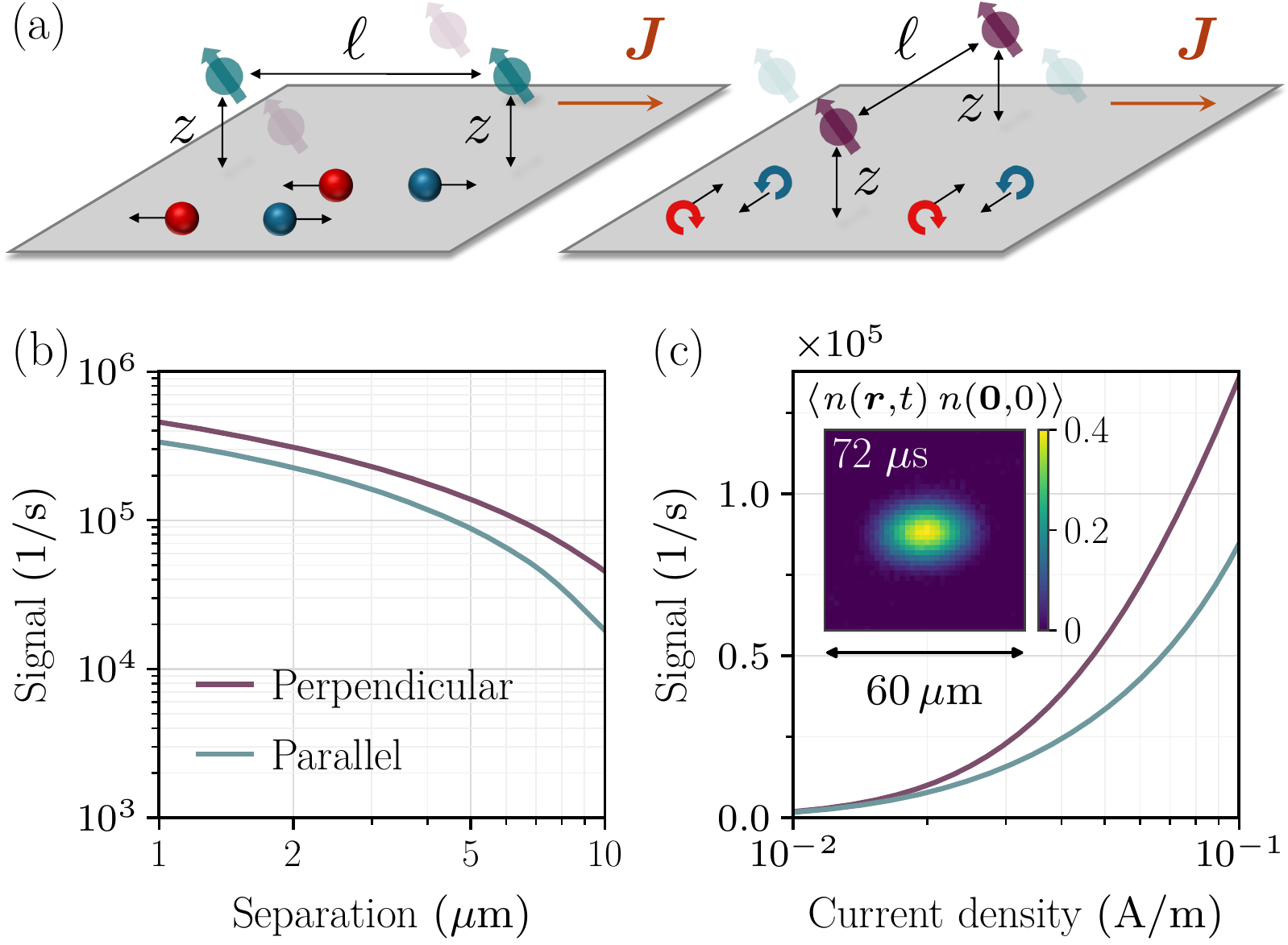}
    \caption{Anisotropic response of charge carriers to an applied current and the resulting noise covariance. (a)~Electric carriers, electrons (blue) or holes (red), drift parallel to $\boldsymbol{J}$, enhancing the noise covariance for NV pairs (teal) aligned along $\boldsymbol{J}$ (left); magnetic vortices feel a Magnus force perpendicular to $\boldsymbol{J}$, enhancing the covariance for perpendicular pairs of NVs (purple, right). Each NV is positioned at a depth $z$ above the sample, with in-plane separation $\boldsymbol\ell$. (b)~Noise covariance $\Gamma(\boldsymbol\ell, z)$ at $z$\,$=$\,$100\,$nm and $\tau$\,$=$\,$100\,\mu$s, versus $\ell$, for perpendicular and parallel NV pairs. (c)~The same versus current density $J$ at $\ell = 5\,\mu$m. Inset: the vortex-charge correlator $\langle n(\boldsymbol r,t)\, n(0,0)\rangle$ at a representative time, $t = 72\,\mu$s ($T = 0.86\, T_c$, $J = 0.1\,$A/m), as calculated using a two-flavor stochastic exclusion process; the anisotropic Gaussian form is characteristic of diffusive vortex hydrodynamics.}
    \label{fig:setup}
\end{figure}

In the present work, we address an aspect of nonequilibrium dynamics that is relatively underexplored, namely the \emph{spatial} structure of nonequilibrium noise~\cite{Zhang2024}, which can be measured using covariance magnetometry~\cite{Rovny2022, Rovny2024, 7xlj-52xt, hosseinabadi2026theory}. This analysis is inspired by the suggestion that (some) strange metals are ``failed superconductors''~\cite{KapitulnikKivelsonSpivak2019, phillips2003elusive, yang2022signatures} (based on evidence, e.g., from the Nernst effect~\cite{PhysRevB.73.024510}). According to this class of models, transport in strange metals is dissipative because of vortex motion. Our main conclusion is that vortex-based transport leads to qualitatively different spatial noise correlations than quasiparticle-based transport (previously studied in Ref.~\cite{Zhang2024}). To address this question in a tractable context, we consider the simplest example of a ``vortex metal,'' namely a two-dimensional superconductor close to its Berezinskii--Kosterlitz--Thouless (BKT) transition~\cite{Berezinskii1971,Kosterlitz1973,Kosterlitz1974}. 
Above the BKT transition, free vortices drift under an applied current, giving rise to a finite linear-response conductivity~\cite{HalperinNelson1979,ambegaokar1980dynamics,BeasleyMooijOrlando1979,Minnhagen1987,HebardFiory1980,FioryHebard1983,Saito2017review,Ugeda2016,Ge2015}. Near (but below) the transition, a finite applied current leads to an intrinsically nonlinear dissipative response, through vortex unbinding. In this latter regime, we develop a quantitative theory of covariance magnetometry, which illustrates the general point that vortices have sharp noise signatures (Fig.~\ref{fig:setup}). As we will show, these signatures are large enough to be detectable with current experimental technology using nitrogen-vacancy centers in diamond~\cite{Casola2018,barry2020,Degen2017review,Pelliccione2016,Andersen2019,Ku2020,Vool2021,Jenkins2022,Ariyaratne2018,Kolkowitz2015,LeeWong2020,Zhou2021, Agarwal2017,RodriguezNieva2018,Machado2023,Chatterjee2019,Dolgirev2022,Rovny2024}. Even in the context of thin-film superconductors, covariance magnetometry can yield unambiguous evidence that the resistive transition is genuinely vortex-driven, rather than dominated by Cooper-pair breaking or sample disorder. 

The key observation underlying our analysis is that the electric and magnetic charge carriers leave a distinctive directional fingerprint in the spatial correlations of the noise produced above the sample. Under an applied in-plane current $\boldsymbol{J}$, electric quasiparticles (electrons or holes) drift parallel to it, whereas magnetic vortices feel a Magnus force and drift perpendicular to it. Consequently, current fluctuations are temporally correlated along the direction of carrier motion, and this directional structure can be revealed by measuring the noise covariance at two nearby points.

The protocol we propose is illustrated in Fig.~\ref{fig:setup}(a). The sample is held at a temperature slightly below $T_{\rm BKT}$ and biased with a small current. Two NV centers, separated by a distance $\ell$ comparable to or larger than the typical vortex separation, monitor the magnetic noise. We compare the noise covariance for NV pairs aligned {parallel} versus {perpendicular} to $\boldsymbol{J}$. If the transition is driven by Cooper-pair breaking, the parallel configuration shows an enhanced covariance; if it is driven by vortex proliferation, the perpendicular one does. Notably, this protocol provides \emph{robust} signatures of the BKT transition: since the underlying signal is anisotropic, it is naturally distinguished from any isotropic noise enhancement due to Joule heating, confirming that the measured signal is genuinely nonequilibrium and topological in origin. Beyond identifying the transition, it also probes the nonequilibrium dynamics of the vortex system. Since vortex charge is conserved, its fluctuations relax on much longer time and length scales than nonconserved quantities. This conservation law leads to large-scale {fluctuating hydrodynamics} of vortices, directly accessible via covariance magnetometry, and provides a new platform to study nonequilibrium hydrodynamics with spatially and temporally resolved measurements. Thus, our proposal is complementary to those addressing the critical properties of the BKT transition~\cite{Dolgirev2022,Curtis2024BKT}; our interest is in identifying signatures of vortex-based transport, rather than these critical properties themselves.

To make this proposal quantitative, we develop a theoretical framework that combines (i)~a renormalization-group (RG) treatment of vortex unbinding and an estimate of the field-driven generation rate following the Ambegaokar--Halperin--Nelson--Siggia (AHNS) approach~\cite{ambegaokar1980dynamics}; (ii)~a two-flavor stochastic exclusion process (SEP) that captures the hydrodynamics of charged vortices on mesoscopic scales; and (iii)~a magnetostatic translation of the resulting density correlations into measurable NV phase covariances. Owing to a phase-space tradeoff, the predicted signal depends only weakly on the NV depth in the regime $z \ll \ell$, making our proposal robust to the sub-100\,nm depth fluctuations encountered in current implantation protocols.

\textit{Theoretical framework.}---We work in the genuinely 2D limit, where the Pearl length~\cite{Pearl1964} of the film exceeds all other length scales of interest---a regime well realized in ultrathin films~\cite{Saito2017review}, monolayer transition-metal dichalcogenides~\cite{Ugeda2016}, and many cuprate-derived 2D superconductors. The diffraction limit on optical addressing of individual NV centers~\cite{Rovny2022} sets a minimum NV separation of $\ell_{\min} \sim 500$\,nm, well above the microscopic vortex core ($\xi \sim 1$--$10$\,nm) and the typical pair size ($\sim 10$--$100$\,nm); however, vortex hydrodynamics governs the physics on much larger scales, so the noise covariance remains robust on the order of microns. In practice, NV centers are randomly implanted at low density to guarantee pairs with suitable separations along both directions and to enable ensemble averaging over many pairs.

An applied current $\boldsymbol J$ exerts a Magnus force $\boldsymbol F = \Phi_0\, \boldsymbol J\times\hat z$ on each vortex. This perturbation tilts the logarithmic vortex--vortex interaction $$U(\boldsymbol r) = 2 q_0^2 \int_{a_0}^{|\boldsymbol r|} dr'/[\epsilon(\ell')\,r'] - q^{}_0\,\boldsymbol F\cdot\boldsymbol r,$$ where $\epsilon(\ell') = K_0/K(\ell')$ is the scale-dependent dielectric constant of the vortex Coulomb gas, $a_0$ is a microscopic UV cutoff, and $\ell' = \ln(r'/a_0)$. Here, $K(\ell')$ is the running dimensionless superfluid stiffness with bare value $K_0\equiv K(0)$, following from the standard BKT RG flow~\cite{Kosterlitz1974,ambegaokar1980dynamics}. The tilt creates a saddle point in $U$ at a critical separation $r_c$ defined by $F r_c\, \epsilon(\ell_c) = 2 q_0$, with $\ell_c \equiv \ln(r_c/a_0)$. At this scale, the vortices unbind, and the current effectively cuts off this RG flow. Following the Fokker--Planck saddle-point analysis of Ref.~\cite{ambegaokar1980dynamics}, the vortex generation rate per unit area becomes
\begin{equation}
    R = 2D\, \frac{y(\ell_c)^2}{r_c^4}\, \exp\!\big(2\pi K(\ell_c)\big),
    \label{eq:R}
\end{equation}
where $y(\ell_c)$ is the renormalized running vortex fugacity at $\ell_c$ and $D$ is the vortex diffusion constant. The corresponding free-vortex density is $n_f = [R/(2\pi D K(\ell_c))]^{1/2}$, which produces the nonlinear $I$-$V$ relation
\begin{equation}
    \boldsymbol E = \mu_v\, \Phi_0^2\, n_f\, \boldsymbol J,
    \label{eq:IV}
\end{equation}
with $\mu_v = D/k_B T$ the vortex mobility. Equation~\eqref{eq:IV} reproduces the celebrated nonlinear $I$-$V$ characteristics of BKT systems and provides a means to fix the otherwise free RG parameters by matching to transport data. Detailed derivations are presented in the Supplemental Material (SM)~\footnote{See Supplemental Material for detailed derivations of the BKT renormalization-group flow, the Fokker--Planck saddle-point analysis of the vortex generation rate, the SEP calibration, and the computation of the vortex correlator.}.

\begin{figure}[t]
    \centering
    \includegraphics[width=\linewidth]{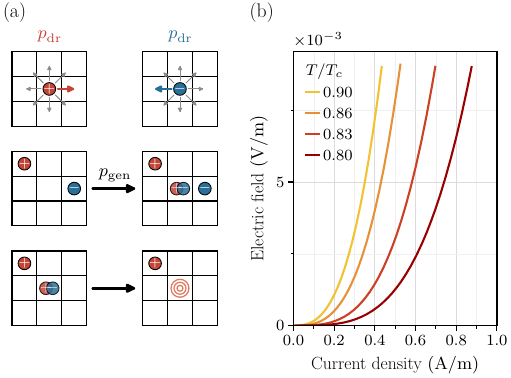}
    \caption{(a)~Update rules of the two-flavor SEP. With probability $p_{\rm dr}$, particles drift along the direction set by their charge ($+\hat x$ for $+$, $-\hat x$ for $-$); otherwise they hop to one of the eight nearest neighbors uniformly. Hopping is rejected if the target site is occupied by a particle of the \emph{same} charge but allowed if it hosts one of \emph{opposite} charge. With probability $p_{\rm gen}$, an empty site nucleates a $\pm$ vortex pair. Two opposite charges on the same site deterministically annihilate. (b)~Nonlinear $I$-$V$ curve from Eq.~\eqref{eq:IV} for a Nb thin film at $T = 0.8, 0.83, 0.86, 0.9\, T_c$.}
    \label{fig:sep}
\end{figure}

To compute spatially resolved noise correlations, we need not just the mean vortex density but also its space-time correlation function. Solving the full Fokker--Planck equation for two interacting vortex species is intractable, so we adopt a coarse-grained mesoscopic description: a two-flavor SEP capturing the leading processes---driven drift, diffusion, hard-core repulsion within each species, and pair annihilation---that govern the underlying vortex dynamics. The model lives on a 2D lattice of spacing $dL$, with each site permitted to host at most one positive ($+$) and at most one negative ($-$) vortex. The discrete-time update [Fig.~\ref{fig:sep}(a)] consists of: (i)~\emph{generation}, where every empty site nucleates a $\pm$ vortex pair with probability $p_{\rm gen}$; (ii)~\emph{biased hopping}, where with probability $p_{\rm dr}$ each particle attempts a drift along $\pm \hat x$ (positive in $+\hat x$, negative in $-\hat x$) and otherwise attempts a uniform move to one of the eight neighboring sites, with moves rejected if the target hosts a same-charge particle; and (iii)~\emph{annihilation}, where a vortex and an antivortex are mutually removed if they land on the same site. One simulation time step corresponds to physical time $dt$. The four parameters $\{p_{\rm gen}, p_{\rm dr}, dL, dt\}$ are fixed from the continuum description (see SM); in particular, $p_{\rm gen} = R\, dL^2\, dt$ uses the AHNS rate from Eq.~\eqref{eq:R}.

We evolve the SEP for a sufficiently long time to reach a steady state and measure the vortex-charge correlator $\langle n(\boldsymbol r,t)\, n(0,0)\rangle$ of $n = n_+ - n_-$ by averaging over space-time translations. Each vortex carries flux $\Phi_0$ perpendicular to the film, and the field above the surface decays evanescently as a solution of Laplace's equation. The magnetic-field correlator at the NV depth $z$ is therefore (see SM for full derivation)
\begin{equation}\label{eq:bfield}
        \langle B(\boldsymbol{r},z,t) B(0,z,0) \rangle = 
        \frac{\Phi_0^2}{dL^4} \frac{1}{L^2 \mathcal{T}} \sum_{\boldsymbol{k},\omega} 
        S_n(\boldsymbol k,\omega)
         e^{i\boldsymbol{k}\cdot \boldsymbol{r} -2k z+i\omega t},
\end{equation}
with $S_n(\boldsymbol k,\omega)$\,$\equiv$\,$\langle n(\boldsymbol{k},\omega) n(-\boldsymbol{k},-\omega) \rangle$ the Fourier transform of the vortex-density correlator, $L^2$ the size of the simulated lattice, and $\mathcal{T}$ the total simulation time. From Eq.~\eqref{eq:bfield}, it is apparent that the depth $z$ acts as a low-pass spatial filter with cutoff $|\boldsymbol k|\sim 1/(2z)$. The phase covariance accumulated by two NVs with an intrinsic lifetime $\tau$ is then
\begin{equation}
    \Gamma(\boldsymbol r, z;\tau) = \gamma^2\!\int_0^{\infty}\! dt\,e^{-t/\tau}\langle B(\boldsymbol r, z, t)\, B(0, z, 0)\rangle,
    \label{eq:Gamma}
\end{equation}
where $\gamma = 2\pi\times 28\,$GHz/T is the NV gyromagnetic ratio. Since the relevant noise frequencies for vortex hydrodynamics lie well below the NV electron-spin resonance ($\sim$\,GHz), the noise spectrum is essentially white over the bandwidth of interest, and dynamical-decoupling sequences~\cite{Rovny2022} simply integrate the noise over $\tau$. Effectively, $\tau$ sets an interrogation window which is bounded by the intrinsic NV coherence time, $\tau \le T_2$. We report $\Gamma(\boldsymbol r, z; \tau)$ as the central observable.

\textit{Results.}---We illustrate the framework with parameters representative of a Nb thin film, a well-studied 2D superconductor exhibiting BKT phenomenology~\cite{Saito2017review}: $T_c = 7.3\,$K, upper critical field $B_{c2} = 5\,$T, and normal-state sheet resistance $\rho_N = 7\,\Omega/\square$. The vortex core size is set by the Ginzburg--Landau coherence length $\xi = \sqrt{\Phi_0/(2\pi B_{c2})} \approx 8\,$nm, which we identify with the UV cutoff $a_0$. The vortex diffusion constant is given by the Bardeen--Stephen relation $D = \rho_N\, k_B T/(\Phi_0\, B_{c2})$~\cite{BardeenStephen1965,Tinkham2004}. The bare RG inputs are $K_0 = (2/\pi)(T_c/T)$ and a fugacity $y_0 = 0.02$, chosen to give a nonlinear $I$-$V$ curve consistent with thin-film experiments~\cite{HebardFiory1980,FioryHebard1983}; none of the qualitative results below depend on the precise value of $y_0$. Figure~\ref{fig:sep}(b) shows the resulting nonlinear $I$-$V$ characteristics at four temperatures, with a power-law exponent $\sim 3$ near $T_c$ in agreement with classic BKT-film experiments~\cite{HebardFiory1980,FioryHebard1983,Minnhagen1987}.

Feeding the RG-derived rate $R$ and the corresponding $J$ into the SEP and evolving to steady state, we compute $\langle n(\boldsymbol r,t)\, n(0,0)\rangle$ at $T = 0.86\, T_c$ and $J = 0.1\,$A/m. As shown in the inset of Fig.~\ref{fig:setup}(c), this correlator takes the form of a single anisotropic Gaussian that broadens diffusively in time, with a noticeable elongation along the current direction. This is the spatial signature of vortex hydrodynamics with an effectively anisotropic diffusion tensor: in the lab frame, vortices of opposite sign are advected in opposite directions perpendicular to $\boldsymbol J$, smearing $\langle n \,n\rangle$ along that axis.

From the vortex correlator we compute, via Eqs.~\eqref{eq:bfield}--\eqref{eq:Gamma}, the noise covariance as a function of the NV separation, for NV pairs aligned perpendicular (purple) or parallel (teal) to the current. Across more than a decade in $\ell$ [Fig.~\ref{fig:setup}(b)], the perpendicular-aligned pair shows a robust $\sim 1.5$--$2\times$ enhancement over the parallel-aligned pair. The absolute scale of $\Gamma$ is in the $10^4$--$10^5\,$s$^{-1}$ range, comparable to $1/T_2$ for typical shallow NVs---precisely the regime in which the covariance contributes a measurable phase contrast in dynamical-decoupling sequences. Fixing $\ell = 5\,\mu$m and varying $J$ [Fig.~\ref{fig:setup}(c)], we find that the anisotropy persists across the entire current range, confirming that it originates in the directional structure of vortex hydrodynamics rather than fine-tuning of parameters. The total noise grows with $J$ due to an increasing vortex density, but the anisotropy ratio remains roughly constant.

\begin{figure}[t]
    \centering
    \includegraphics[width=\linewidth]{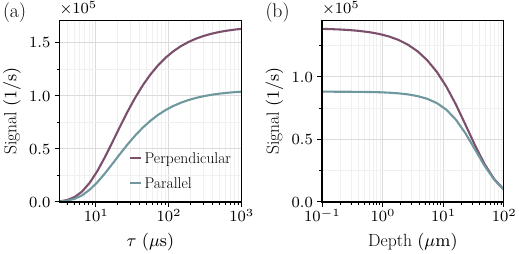}
    \caption{Robustness of the predicted signal at fixed $\ell = 5\,\mu$m. (a)~$\Gamma$ versus the intrinsic NV lifetime $\tau$, at $z = 100\,$nm. The signal saturates for $\tau\gtrsim 100\,\mu$s, set by the noise correlation time, with the parallel/perpendicular anisotropy roughly constant across the experimentally relevant range. (b)~$\Gamma$ versus depth $z$ at $\tau = 100\,\mu$s. The signal is approximately depth-independent in the regime $z\ll \ell$, owing to a phase-space tradeoff (see text).}
    \label{fig:sensitivity}
\end{figure}

\textit{Robustness.}---A practical advantage of covariance magnetometry on extended 2D samples is that the signal is particularly forgiving of two key experimental imperfections: limited NV coherence and depth fluctuations of the implanted NVs. We first vary the lifetime $\tau$ at fixed $\ell = 5\,\mu$m and $z = 100\,$nm [Fig.~\ref{fig:sensitivity}(a)]. The covariance grows with $\tau$ and saturates around $\tau \sim 100\,\mu$s, set by the correlation time of the magnetic noise; beyond this scale, additional integration time does not yield further accumulation of phase contrast. For shallow NVs with $T_2$ in the $10$--$100\,\mu$s range~\cite{Bluvstein2019,Sangtawesin2019,barry2020}, the accessible $\Gamma$ lies between $10^4$ and $10^5\,$s$^{-1}$, well within reach of state-of-the-art covariance protocols~\cite{Rovny2022}. Crucially, the fractional anisotropy is roughly $\tau$-independent in this range, so working with a more readily achievable coherence does not compromise the directional signature.

Next, we vary the NV depth at fixed $\ell$ [Fig.~\ref{fig:sensitivity}(b)]. Strikingly, the noise covariance decreases with $z$ but only weakly when $z < \ell$. This is a generic feature of correlated 2D noise sensing: the depth $z$ acts as a low-pass spatial filter with bandwidth $\sim 1/z$ via the $e^{-2|\boldsymbol k| z}$ factor in Eq.~\eqref{eq:bfield}, while the separation $\ell$ acts as a band-pass filter with bandwidth $\sim 1/\ell$ via $e^{i\boldsymbol k\cdot \boldsymbol\ell}$. When $z \ll \ell$, the latter is the more restrictive cut, and the covariance becomes insensitive to $z$~\cite{Zhang2024}. As a consequence, depth fluctuations in the implant process do not significantly affect the predicted signal, removing one of the main sources of uncertainty in extracting it. NV centers also have shorter coherence times near the surface, so it is advantageous to use a sufficiently large depth to avoid surface-induced decoherence; choosing $z \sim 100\,$nm yields surface-limited coherence times $T_2 \sim 10\text{--}100\,\mu$s, which suffice to resolve the predicted signal.

\textit{Discussion.}---Focusing on the concrete example of thin-film superconductors near the BKT transition, we have proposed and analyzed a protocol that detects the vortex-driven nature of transport via a directional fingerprint that is markedly distinct from any heating-induced isotropic noise enhancement. It is directly applicable to anomalous metallic phases of 2D superconductors~\cite{KapitulnikKivelsonSpivak2019}; measurement of perpendicular-aligned correlated noise would provide strong evidence that the residual resistance is due to vortex motion rather than, e.g., uncondensed electron-like fluctuations. In those settings, the expected magnitude of the signal due to vortices can be estimated from the nonlinear $I$-$V$ curve. More generally, the directional fingerprint described here applies to any 2D system in which charge transport is mediated by topological excitations subject to a transverse driving force. Potential applications include skyrmion~\cite{Muhlbauer2009,Yu2010,NagaosaTokura2013,Jiang2017} and meron~\cite{Yu2018Meron} textures in chiral magnets, and 2D superfluids of dipolar excitons or polaritons.

Our results are based on the simplification (valid in the regime of most interest) that free vortices dominate the observed response, and are generated via a Poisson process. Natural methodological extensions would incorporate the noise signal due to weakly bound vortex dipoles (which should be oriented transverse to the current, and have a size comparable to the crossover scale $r_c$), as well as spatial correlations in the vortex creation process. 
These signals are potentially measurable through higher cumulants, which are accessible by simultaneously reading out three or more NVs, opening a route to the full counting statistics of charge in low-dimensional systems~\cite{Wei2022,Rosenberg2024,Wienand2024,McCulloch2023,Gopalakrishnan2024non,Samajdar2024,Krajnik2022a,Gopalakrishnan2024diff,Krajnik2022b}. With dense NV arrays~\cite{Cambria2025,Cheng2025}, one could go further still and reconstruct the full real-space hydrodynamic correlator $\langle n(\boldsymbol r,t)\,n(0,0)\rangle$ rather than only its anisotropy, providing access to the vortex viscosity, density, and nonlinear corrections.

More broadly, the present work---together with our recent analysis of nonequilibrium current noise in normal metals~\cite{Zhang2024}---establishes covariance magnetometry as a uniquely powerful probe of the spatial structure of mesoscopic fluctuations. Whereas conventional transport measurements integrate over the entire sample and discard directional information, NV covariance protocols access the noise tensor in real space and on micron scales. This opens further experimental and theoretical avenues for correlated quantum sensing as a tool for out-of-equilibrium condensed-matter physics.

\begin{acknowledgments}
\textit{Acknowledgments.}---We thank Kai-Hung Cheng, Nathalie de Leon, Zeeshawn Kazi, Ewan McCulloch, and Dror Orgad for useful discussions. S.G. and Y.Z. acknowledge support from NSF QuSEC-TAQS OSI 2326767. R.S. was supported by the Princeton Quantum Initiative Fellowship.
\end{acknowledgments}

\bibliography{nvnoise}

@article{Berezinskii1971,
  author = {Berezinski\u{\i}, V. L.},
  title = "{Destruction of Long-range Order in One-dimensional and Two-dimensional Systems Possessing a Continuous Symmetry Group. {II}. {Q}uantum Systems}",
  journal = {Sov. Phys. JETP},
  volume = {34},
  pages = {610},
  year = {1972},
  url={https://inspirehep.net/files/0f7b50c47ec26bed99a50ff199960259}
}

@article{Kosterlitz1973,
  author = {Kosterlitz, J. M. and Thouless, D. J.},
  title = {Ordering, metastability and phase transitions in two-dimensional systems},
  journal = {J. Phys. C: Solid State Phys.},
  volume = {6},
  pages = {1181--1203},
  year = {1973},
  doi = {10.1088/0022-3719/6/7/010}
}

@article{Kosterlitz1974,
  author = {Kosterlitz, J. M.},
  title = {The critical properties of the two-dimensional {$xy$} model},
  journal = {J. Phys. C: Solid State Phys.},
  volume = {7},
  pages = {1046--1060},
  year = {1974},
  doi = {10.1088/0022-3719/7/6/005}
}

@article{HalperinNelson1979,
  author = {Halperin, B. I. and Nelson, David R.},
  title = {Resistive transition in superconducting films},
  journal = {J. Low Temp. Phys.},
  volume = {36},
  pages = {599--616},
  year = {1979},
  doi = {10.1007/BF00116988}
}

@article{ambegaokar1980dynamics,
  author = {Ambegaokar, Vinay and Halperin, B. I. and Nelson, David R. and Siggia, Eric D.},
  title = {Dynamics of superfluid films},
  journal = {Phys. Rev. B},
  volume = {21},
  pages = {1806--1826},
  year = {1980},
  doi = {10.1103/PhysRevB.21.1806}
}

@article{BeasleyMooijOrlando1979,
  author = {Beasley, M. R. and Mooij, J. E. and Orlando, T. P.},
  title = {Possibility of vortex-antivortex pair dissociation in two-dimensional superconductors},
  journal = {Phys. Rev. Lett.},
  volume = {42},
  pages = {1165--1168},
  year = {1979},
  doi = {10.1103/PhysRevLett.42.1165}
}

@article{Minnhagen1987,
  author = {Minnhagen, Petter},
  title = {The two-dimensional {C}oulomb gas, vortex unbinding, and superfluid-superconducting films},
  journal = {Rev. Mod. Phys.},
  volume = {59},
  pages = {1001--1066},
  year = {1987},
  doi = {10.1103/RevModPhys.59.1001}
}

@article{HebardFiory1980,
  author = {Hebard, A. F. and Fiory, A. T.},
  title = {Evidence for the {Kosterlitz-Thouless} transition in thin superconducting aluminum films},
  journal = {Phys. Rev. Lett.},
  volume = {44},
  pages = {291--294},
  year = {1980},
  doi = {10.1103/PhysRevLett.44.291}
}

@article{FioryHebard1983,
  author = {Fiory, A. T. and Hebard, A. F. and Glaberson, W. I.},
  title = {Superconducting phase transitions in indium/indium-oxide thin-film composites},
  journal = {Phys. Rev. B},
  volume = {28},
  pages = {5075--5087},
  year = {1983},
  doi = {10.1103/PhysRevB.28.5075}
}

@article{Pearl1964,
  author = {Pearl, J.},
  title = {Current distribution in superconducting films carrying quantized fluxoids},
  journal = {Appl. Phys. Lett.},
  volume = {5},
  pages = {65--66},
  year = {1964},
  doi = {10.1063/1.1754056}
}

@article{Saito2017review,
  author = {Saito, Yu and Nojima, Tsutomu and Iwasa, Yoshihiro},
  title = {Highly crystalline 2{D} superconductors},
  journal = {Nat. Rev. Mater.},
  volume = {2},
  pages = {16094},
  year = {2017},
  doi = {10.1038/natrevmats.2016.94}
}

@article{Ugeda2016,
  author = {Ugeda, Miguel M. and Bradley, Aaron J. and Zhang, Yi and Onishi, Seita and Chen, Yi and Ruan, Wei and Ojeda-Aristizabal, Claudia and Ryu, Hyejin and Edmonds, Mark T. and Tsai, Hsin-Zon and Riss, Alexander and Mo, Sung-Kwan and Lee, Dunghai and Zettl, Alex and Hussain, Zahid and Shen, Zhi-Xun and Crommie, Michael F.},
  title = {Characterization of collective ground states in single-layer {NbSe$_2$}},
  journal = {Nat. Phys.},
  volume = {12},
  pages = {92--97},
  year = {2016},
  doi = {10.1038/nphys3527}
}

@article{Ge2015,
  author = {Ge, J.-F. and Liu, Z.-L. and Liu, C. and Gao, C.-L. and Qian, D. and Xue, Q.-K. and Liu, Y. and Jia, J.-F.},
  title = {Superconductivity above 100 {K} in single-layer {FeSe} films on doped {SrTiO}$_3$},
  journal = {Nat. Mater.},
  volume = {14},
  pages = {285--289},
  year = {2015},
  doi = {10.1038/nmat4153}
}

@article{BardeenStephen1965,
  author = {Bardeen, John and Stephen, M. J.},
  title = {Theory of the motion of vortices in superconductors},
  journal = {Phys. Rev.},
  volume = {140},
  pages = {A1197--A1207},
  year = {1965},
  doi = {10.1103/PhysRev.140.A1197}
}

@book{Tinkham2004,
  author = {Tinkham, Michael},
  title = {Introduction to Superconductivity},
  edition = {2},
  publisher = {Dover},
  year = {2004}
}

@article{sachdev2025foot,
  title="{The foot, the fan, and the cuprate phase diagram: Fermi-volume-changing quantum phase transitions}",
  author={Sachdev, Subir},
  journal={Physica C: Supercond. Appl.},
  volume={633},
  pages={1354707},
  year={2025},
  publisher={Elsevier},
  doi={10.1016/j.physc.2025.1354707}
}

@book{nozieres2018theory,
  title="{Theory of interacting Fermi systems}",
  author={Nozières, Philippe},
  year={2018},
  publisher={CRC Press},
  doi={10.1201/9780429495724}
}

@article{KapitulnikKivelsonSpivak2019,
  author = {Kapitulnik, Aharon and Kivelson, Steven A. and Spivak, Boris},
  title = {{C}olloquium: Anomalous metals: Failed superconductors},
  journal = {Rev. Mod. Phys.},
  volume = {91},
  pages = {011002},
  year = {2019},
  doi = {10.1103/RevModPhys.91.011002}
}

@article{phillips2003elusive,
  title="{The Elusive Bose Metal}",
  author={Phillips, Philip and Dalidovich, Denis},
  journal={Science},
  volume={302},
  number={5643},
  pages={243--247},
  year={2003},
  publisher={American Association for the Advancement of Science},
  doi={10.1126/science.1088253}
}

@article{yang2022signatures,
	author = {Yang, Chao and Liu, Haiwen and Liu, Yi and Wang, Jiandong and Qiu, Dong and Wang, Sishuang and Wang, Yang and He, Qianmei and Li, Xiuli and Li, Peng and Tang, Yue and Wang, Jian and Xie, X. C. and Valles, James M. and Xiong, Jie and Li, Yanrong},
	date = {2022/01/01},
	doi = {10.1038/s41586-021-04239-y},
	id = {Yang2022},
	isbn = {1476-4687},
	journal = {Nature},
	number = {7892},
	pages = {205--210},
	title = {Signatures of a strange metal in a bosonic system},
	url = {https://doi.org/10.1038/s41586-021-04239-y},
	volume = {601},
	year = {2022}}

@article{sachdev2009quantum,
  title={Quantum criticality and black holes},
  author={Sachdev, Subir and M{\"u}ller, Markus},
  journal={J. Phys.: Condens. Matter},
  volume={21},
  number={16},
  pages={164216},
  year={2009},
  doi={10.1088/0953-8984/21/16/164216}
}

@article{PhysRevB.73.024510,
  title = {Nernst effect in high-${T}_{c}$ superconductors},
  author = {Wang, Yayu and Li, Lu and Ong, N. P.},
  journal = {Phys. Rev. B},
  volume = {73},
  issue = {2},
  pages = {024510},
  numpages = {20},
  year = {2006},
  month = {Jan},
  publisher = {American Physical Society},
  doi = {10.1103/PhysRevB.73.024510},
  url = {https://link.aps.org/doi/10.1103/PhysRevB.73.024510}
}

@article{Degen2017review,
  author = {Degen, C. L. and Reinhard, F. and Cappellaro, P.},
  title = {Quantum sensing},
  journal = {Rev. Mod. Phys.},
  volume = {89},
  pages = {035002},
  year = {2017},
  doi = {10.1103/RevModPhys.89.035002}
}

@article{Casola2018,
  author = {Casola, Francesco and van der Sar, Toeno and Yacoby, Amir},
  title = {Probing condensed matter physics with magnetometry based on nitrogen-vacancy centres in diamond},
  journal = {Nat. Rev. Mater.},
  volume = {3},
  pages = {17088},
  year = {2018},
  doi = {10.1038/natrevmats.2017.88}
}

@article{barry2020,
  author = {Barry, John F. and Schloss, Jennifer M. and Bauch, Erik and Turner, Matthew J. and Hart, Connor A. and Pham, Linh M. and Walsworth, Ronald L.},
  title = {Sensitivity optimization for {NV}-diamond magnetometry},
  journal = {Rev. Mod. Phys.},
  volume = {92},
  pages = {015004},
  year = {2020},
  doi = {10.1103/RevModPhys.92.015004}
}

@article{Kolkowitz2015,
  author = {Kolkowitz, S. and Safira, A. and High, A. A. and Devlin, R. C. and Choi, S. and Unterreithmeier, Q. P. and Patterson, D. and Zibrov, A. S. and Manucharyan, V. E. and Park, H. and Lukin, M. D.},
  title = {Probing {Johnson} noise and ballistic transport in normal metals with a single-spin qubit},
  journal = {Science},
  volume = {347},
  pages = {1129--1132},
  year = {2015},
  doi = {10.1126/science.aaa4298}
}

@article{Pelliccione2016,
  author = {Pelliccione, Matthew and Jenkins, Alec and Ovartchaiyapong, Preeti and Reetz, Christopher and Emmanouilidou, Eve and Ni, Ni and Bleszynski Jayich, Ania C.},
  title = {Scanned probe imaging of nanoscale magnetism at cryogenic temperatures with a single-spin quantum sensor},
  journal = {Nat. Nanotechnol.},
  volume = {11},
  pages = {700--705},
  year = {2016},
  doi = {10.1038/nnano.2016.68}
}

@article{Agarwal2017,
  author = {Agarwal, Kartiek and Schmidt, Richard and Halperin, Bertrand and Oganesyan, Vadim and Zar\'and, Gergely and Lukin, Mikhail D. and Demler, Eugene},
  title = {Magnetic noise spectroscopy as a probe of local electronic correlations in two-dimensional systems},
  journal = {Phys. Rev. B},
  volume = {95},
  pages = {155107},
  year = {2017},
  doi = {10.1103/PhysRevB.95.155107}
}

@article{RodriguezNieva2018,
  author = {Rodriguez-Nieva, J. F. and Agarwal, Kartiek and Giamarchi, Thierry and Halperin, B. I. and Lukin, M. D. and Demler, Eugene},
  title = {Probing one-dimensional systems via noise magnetometry with single spin qubits},
  journal = {Phys. Rev. B},
  volume = {98},
  pages = {195433},
  year = {2018},
  doi = {10.1103/PhysRevB.98.195433}
}

@article{Machado2023,
  author = {Machado, Francisco and Demler, Eugene A. and Yao, Norman Y. and Chatterjee, Shubhayu},
  title = {Quantum noise spectroscopy of dynamical critical phenomena},
  journal = {Phys. Rev. Lett.},
  volume = {131},
  pages = {070801},
  year = {2023},
  doi = {10.1103/PhysRevLett.131.070801}
}

@article{Chatterjee2019,
  author = {Chatterjee, Shubhayu and Rodriguez-Nieva, Joaquin F. and Demler, Eugene},
  title = {Diagnosing phases of magnetic insulators via noise magnetometry with spin qubits},
  journal = {Phys. Rev. B},
  volume = {99},
  pages = {104425},
  year = {2019},
  doi = {10.1103/PhysRevB.99.104425}
}

@article{Curtis2024BKT,
  author = {Curtis, Jonathan B. and Maksimovic, Nikola and Poniatowski, Nicholas R. and Yacoby, Amir and Halperin, Bertrand and Narang, Prineha and Demler, Eugene},
  title = {Probing the {Berezinskii-Kosterlitz-Thouless} vortex unbinding transition in two-dimensional superconductors using local noise magnetometry},
  journal = {Phys. Rev. B},
  volume = {110},
  pages = {144518},
  year = {2024},
  doi = {10.1103/PhysRevB.110.144518}
}

@article{Dolgirev2022,
  author = {Dolgirev, Pavel E. and Chatterjee, Shubhayu and Esterlis, Ilya and Zibrov, Alexander A. and Lukin, Mikhail D. and Yao, Norman Y. and Demler, Eugene},
  title = {Characterizing two-dimensional superconductivity via nanoscale noise magnetometry with single-spin qubits},
  journal = {Phys. Rev. B},
  volume = {105},
  pages = {024507},
  year = {2022},
  doi = {10.1103/PhysRevB.105.024507}
}

@article{Rovny2022,
  author = {Rovny, Jared and Yuan, Zhiyang and Fitzpatrick, Mattias and Abdalla, Ahmed I. and Futamura, Levi and Fox, Carter and Cambria, Matthew C. and Kolkowitz, Shimon and de Leon, Nathalie P.},
  title = {Nanoscale covariance magnetometry with diamond quantum sensors},
  journal = {Science},
  volume = {378},
  pages = {1301--1305},
  year = {2022},
  doi = {10.1126/science.ade9858}
}

@article{Rovny2024,
  author = {Rovny, Jared and Gopalakrishnan, Sarang and Bleszynski Jayich, Ania C. and Maletinsky, Patrick and Demler, Eugene and de Leon, Nathalie P.},
  title = {Nanoscale diamond quantum sensors for many-body physics},
  journal = {Nat. Rev. Phys.},
  volume = {6},
  pages = {753--768},
  year = {2024},
  doi = {10.1038/s42254-024-00775-4}
}

@article{hosseinabadi2026theory,
      title="{Theory of Two-Qubit $T_2$ Spectroscopy of Quantum Many-Body Systems}", 
      author={Hossein Hosseinabadi and Pavel E. Dolgirev and Sarang Gopalakrishnan and Amir Yacoby and Eugene Demler and Jamir Marino},
      year={2026},
      eprint={2603.18176},
      archivePrefix={arXiv},
      primaryClass={quant-ph},
      url={https://arxiv.org/abs/2603.18176}, 
      journal={}
}

@article{7xlj-52xt,
  title = {Wideband Covariance Magnetometry below the Diffraction Limit},
  author = {Le, Xuan Hoang and Dolgirev, Pavel E. and Put, Piotr and Peterson, Eric L. and Pillai, Arjun and Zibrov, Alexander A. and Demler, Eugene and Park, Hongkun and Lukin, Mikhail D.},
  journal = {Phys. Rev. Lett.},
  volume = {135},
  issue = {17},
  pages = {170803},
  numpages = {9},
  year = {2025},
  month = {Oct},
  publisher = {American Physical Society},
  doi = {10.1103/7xlj-52xt},
  url = {https://link.aps.org/doi/10.1103/7xlj-52xt}
}

@article{Cambria2025,
  author = {Cambria, Matthew C. and Chand, Saroj and Reiter, Caitlin Mary and Kolkowitz, Shimon},
  title = {Scalable parallel measurement of individual nitrogen-vacancy centers},
  journal = {Phys. Rev. X},
  volume = {15},
  pages = {031015},
  year = {2025},
  doi = {10.1103/PhysRevX.15.031015}
}

@article{Cheng2025,
  author = {Cheng, Kuan-Hao and Kazi, Zaina and Rovny, Jared and Zhang, Bingjie and Nassar, Lila S. and Thompson, Jeff D. and de Leon, Nathalie P.},
  title = {Massively multiplexed nanoscale magnetometry with diamond quantum sensors},
  journal = {Phys. Rev. X},
  volume = {15},
  pages = {031014},
  year = {2025},
  doi = {10.1103/PhysRevX.15.031014}
}

@article{Andersen2019,
  author = {Andersen, Trond I. and Dwyer, Bryan L. and Sanchez-Yamagishi, Javier D. and Rodriguez-Nieva, Joaquin F. and Agarwal, Kartiek and Watanabe, Kenji and Taniguchi, Takashi and Demler, Eugene A. and Kim, Philip and Park, Hongkun and Lukin, Mikhail D.},
  title = {Electron-phonon instability in graphene revealed by global and local noise probes},
  journal = {Science},
  volume = {364},
  pages = {154--157},
  year = {2019},
  doi = {10.1126/science.aaw2104}
}

@article{Ku2020,
  author = {Ku, Mark J. H. and Zhou, Tony X. and Li, Qing and Shin, Young J. and Shi, Jing K. and Burch, Claire and Anderson, Laurel E. and Pierce, Andrew T. and Xie, Yonglong and Hamo, Assaf and Vool, Uri and Zhang, Huiliang and Casola, Francesco and Taniguchi, Takashi and Watanabe, Kenji and Fogler, Michael M. and Kim, Philip and Yacoby, Amir and Walsworth, Ronald L.},
  title = {Imaging viscous flow of the {Dirac} fluid in graphene},
  journal = {Nature},
  volume = {583},
  pages = {537--541},
  year = {2020},
  doi = {10.1038/s41586-020-2507-2}
}

@article{Vool2021,
  author = {Vool, Uri and Hamo, Assaf and Varnavides, Georgios and Wang, Yang and Zhou, Tony X. and Kumar, Nitesh and Dovzhenko, Yuri and Qiu, Ziwei and Garcia, Christina A. C. and Pierce, Andrew T. and Gooth, Johannes and Anikeeva, Polina and Felser, Claudia and Narang, Prineha and Yacoby, Amir},
  title = {Imaging phonon-mediated hydrodynamic flow in {WTe}$_2$},
  journal = {Nat. Phys.},
  volume = {17},
  pages = {1216--1220},
  year = {2021},
  doi = {10.1038/s41567-021-01341-w}
}

@article{Jenkins2022,
  author = {Jenkins, Alec and Baumann, Susanne and Zhou, Haoxin and Meynell, Simon A. and Daipeng, Yang and Watanabe, Kenji and Taniguchi, Takashi and Lucas, Andrew and Young, Andrea F. and Bleszynski Jayich, Ania C.},
  title = {Imaging the breakdown of {O}hmic transport in graphene},
  journal = {Phys. Rev. Lett.},
  volume = {129},
  pages = {087701},
  year = {2022},
  doi = {10.1103/PhysRevLett.129.087701}
}

@article{Ariyaratne2018,
  author = {Ariyaratne, Amila and Bluvstein, Dolev and Myers, Bryan A. and Bleszynski Jayich, Ania C.},
  title = {Nanoscale electrical conductivity imaging using a nitrogen-vacancy center in diamond},
  journal = {Nat. Commun.},
  volume = {9},
  pages = {2406},
  year = {2018},
  doi = {10.1038/s41467-018-04798-1}
}

@article{LeeWong2020,
  author = {Lee-Wong, E. and Xue, R. and Ye, F. and Kreisel, A. and van der Sar, T. and Yacoby, A. and Du, C. R.},
  title = {Nanoscale detection of magnon excitations with variable wavevectors through a quantum spin sensor},
  journal = {Nano Lett.},
  volume = {20},
  pages = {3284--3290},
  year = {2020},
  doi = {10.1021/acs.nanolett.0c00085}
}

@article{Zhou2021,
  author = {Zhou, T. X. and Carmiggelt, J. J. and G\"achter, L. M. and Esterlis, I. and Sels, D. and St\"ohr, R. J. and Du, C. and Fernandez, D. and Rodriguez-Nieva, J. F. and B\"uttner, F. and Demler, E. and Yacoby, A.},
  title = {A magnon scattering platform},
  journal = {Proc. Natl. Acad. Sci. U.S.A.},
  volume = {118},
  pages = {e2019473118},
  year = {2021},
  doi = {10.1073/pnas.2019473118}
}

@article{wang2026reconciling,
title="{Reconciling strange metal transport in CeCoIn$_5$ through the difference of optical and cyclotron effective masses}", 
      author={Jingyuan Wang and Zhenisbek Tagay and Liyu Shi and Jiahao Liang and Nghiep Khoan Duong and Yi Wu and P. M. T. Vianez and F. Ronning and D. G. Rickel and Darrell G. Schlom and K. M. Shen and S. A. Crooker and N. P. Armitage},
      year={2026},
      eprint={2603.23740},
      archivePrefix={arXiv},
      primaryClass={cond-mat.str-el},
      url={https://arxiv.org/abs/2603.23740}, 
      journal={}
}

@article{Zhang2024,
      title={Nanoscale sensing of spatial correlations in nonequilibrium current noise}, 
      author={Yifan Zhang and Rhine Samajdar and Sarang Gopalakrishnan},
      year={2024},
      eprint={2404.15398},
      archivePrefix={arXiv},
      primaryClass={cond-mat.mes-hall},
      url={https://arxiv.org/abs/2404.15398}, 
      journal={}
}

@article{zhang2025evidence,
   title="{Evidence for Bose liquid from anomalous shot noise in nanojunctions of bad metal beta-Ta}", 
      author={Yiou Zhang and Chendi Xie and John Bacsa and Yao Wang and Sergei Urazhdin},
      year={2025},
      eprint={2506.09973},
      archivePrefix={arXiv},
      primaryClass={cond-mat.mtrl-sci},
      url={https://arxiv.org/abs/2506.09973}, 
      journal={}
}

@article{Chen2023,
  author = {Chen, Liyang and Lowder, Daniel T. and Bakali, Erol and Andrews, Aaron M. and Schrenk, Werner and Waas, Monika and Svagera, Robert and Eguchi, Gaku and Prochaska, Lukas and Wang, Yiming and Setty, Chandan and Sur, Shouvik and Si, Qimiao and Paschen, Silke and Natelson, Douglas},
  title = {Shot noise in a strange metal},
  journal = {Science},
  volume = {382},
  pages = {907--911},
  year = {2023},
  doi = {10.1126/science.abq6100}
}

@article{Wang2022shot,
 title = "{Shot noise and universal Fano factor as a characterization of strongly correlated metals}",
  author = {Wang, Yiming and Setty, Chandan and Sur, Shouvik and Chen, Liyang and Paschen, Silke and Natelson, Douglas and Si, Qimiao},
  journal = {Phys. Rev. Res.},
  volume = {6},
  issue = {4},
  pages = {L042045},
  numpages = {5},
  year = {2024},
  month = {Nov},
  publisher = {American Physical Society},
  doi = {10.1103/PhysRevResearch.6.L042045},
  url = {https://link.aps.org/doi/10.1103/PhysRevResearch.6.L042045}
}

@article{Nikolaenko2023,
  author = {Nikolaenko, Andrey and Sachdev, Subir and Patel, Aavishkar A.},
  title = {Theory of shot noise in strange metals},
  journal = {Phys. Rev. Res.},
  volume = {5},
  pages = {043143},
  year = {2023},
  doi = {10.1103/PhysRevResearch.5.043143}
}

@article{WuFoster2024,
  author = {Wu, Tsz Chun and Foster, Matthew S.},
  title = {Suppression of shot noise in a dirty marginal {Fermi} liquid},
  journal = {Phys. Rev. B},
  volume = {110},
  pages = {L081102},
  year = {2024},
  doi = {10.1103/PhysRevB.110.L081102}
}

@article{Green2006,
  author = {Green, A. G. and Moore, J. E. and Sondhi, S. L. and Vishwanath, A.},
  title = {Current noise in the vicinity of the {2D} superconductor-insulator quantum critical point},
  journal = {Phys. Rev. Lett.},
  volume = {97},
  pages = {227003},
  year = {2006},
  doi = {10.1103/PhysRevLett.97.227003}
}

@article{McCulloch2023,
  author = {McCulloch, Ewan and De Nardis, Jacopo and Gopalakrishnan, Sarang and Vasseur, Romain},
  title = {Full counting statistics of charge in chaotic many-body quantum systems},
  journal = {Phys. Rev. Lett.},
  volume = {131},
  pages = {210402},
  year = {2023},
  doi = {10.1103/PhysRevLett.131.210402}
}

@article{Wei2022,
  author = {Wei, David and Rubio-Abadal, Antonio and Ye, Bingtian and Machado, Francisco and Kemp, Jack and Srakaew, Kritsana and Hollerith, Simon and Rui, Jun and Gopalakrishnan, Sarang and Yao, Norman Y. and Bloch, Immanuel and Zeiher, Johannes},
  title = {Quantum gas microscopy of {Kardar-Parisi-Zhang} superdiffusion},
  journal = {Science},
  volume = {376},
  pages = {716--720},
  year = {2022},
  doi = {10.1126/science.abk2397}
}

@article{Wienand2024,
  author = {Wienand, Julian F. and Karch, Simon and Impertro, Alexander and Schweizer, Christian and McCulloch, Ewan and Vasseur, Romain and Gopalakrishnan, Sarang and Aidelsburger, Monika and Bloch, Immanuel},
  title = {Emergence of fluctuating hydrodynamics in chaotic quantum systems},
  journal = {Nat. Phys.},
  volume = {20},
  pages = {1732--1737},
  year = {2024},
  doi = {10.1038/s41567-024-02611-z}
}

@article{Rosenberg2024,
  author = {Rosenberg, E. and Andersen, T. I. and Samajdar, R. and Petukhov, A. and Hoke, J. C. and Abanin, D. and Bengtsson, A. and Drozdov, I. K. and Erickson, C. and Klimov, P. V. and others},
  title = {Dynamics of magnetization at infinite temperature in a {Heisenberg} spin chain},
  journal = {Science},
  volume = {384},
  pages = {48--53},
  year = {2024},
  doi = {10.1126/science.adi7877}
}

@article{Gopalakrishnan2024non,
  author = {Gopalakrishnan, Sarang and McCulloch, Ewan and Vasseur, Romain},
  title = {Non-{G}aussian diffusive fluctuations in {D}irac fluids},
  journal = {Proc. Natl. Acad. Sci. U.S.A.},
  volume = {121},
  pages = {e2403327121},
  year = {2024},
  doi = {10.1073/pnas.2403327121}
}

@article{Samajdar2024,
  author = {Samajdar, Rhine and McCulloch, Ewan and Khemani, Vedika and Vasseur, Romain and Gopalakrishnan, Sarang},
  title = {Quantum turnstiles for robust measurement of full counting statistics},
  journal = {Phys. Rev. Lett.},
  volume = {133},
  pages = {240403},
  year = {2024},
  doi = {10.1103/PhysRevLett.133.240403}
}

@article{Krajnik2022a,
  author = {Krajnik, {\v{Z}}iga and Ilievski, Enej and Prosen, Toma{\v{z}}},
  title = {Absence of normal fluctuations in an integrable magnet},
  journal = {Phys. Rev. Lett.},
  volume = {128},
  pages = {090604},
  year = {2022},
  doi = {10.1103/PhysRevLett.128.090604}
}

@article{Gopalakrishnan2024diff,
  author = {Gopalakrishnan, Sarang and Morningstar, Alan and Vasseur, Romain and Khemani, Vedika},
  title = {Distinct universality classes of diffusive transport from full counting statistics},
  journal = {Phys. Rev. B},
  volume = {109},
  pages = {024417},
  year = {2024},
  doi = {10.1103/PhysRevB.109.024417}
}

@article{Krajnik2022b,
  author = {Krajnik, {\v{Z}}iga and Schmidt, Johannes and Pasquier, Vincent and Ilievski, Enej and Prosen, Toma{\v{z}}},
  title = {Exact anomalous current fluctuations in a deterministic interacting model},
  journal = {Phys. Rev. Lett.},
  volume = {128},
  pages = {160601},
  year = {2022},
  doi = {10.1103/PhysRevLett.128.160601}
}

@article{Bluvstein2019,
  author = {Bluvstein, Dolev and Zhang, Zhiran and McLellan, Claire A. and Williams, Nicolas R. and Bleszynski Jayich, Ania C.},
  title = {Extending the quantum coherence of a near-surface qubit by coherently driving the paramagnetic surface environment},
  journal = {Phys. Rev. Lett.},
  volume = {123},
  pages = {146804},
  year = {2019},
  doi = {10.1103/PhysRevLett.123.146804}
}

@article{Sangtawesin2019,
  author = {Sangtawesin, Sorawis and Dwyer, Bryan L. and Srinivasan, Srikanth and Allred, James J. and Rodgers, Lila V. P. and de Greve, Kristiaan and Stacey, Alastair and Dontschuk, Nikolai and O'Donnell, Kane M. and Hu, Di and Evans, David A. and Jaye, Cherno and Fischer, Daniel A. and Markham, Matthew L. and Twitchen, Daniel J. and Park, Hongkun and Lukin, Mikhail D. and de Leon, Nathalie P.},
  title = {Origins of diamond surface noise probed by correlating single-spin measurements with surface spectroscopy},
  journal = {Phys. Rev. X},
  volume = {9},
  pages = {031052},
  year = {2019},
  doi = {10.1103/PhysRevX.9.031052}
}

@article{Muhlbauer2009,
  author  = {M{\"u}hlbauer, S. and Binz, B. and Jonietz, F. and Pfleiderer, C. and Rosch, A. and Neubauer, A. and Georgii, R. and B{\"o}ni, P.},
  title   = {Skyrmion lattice in a chiral magnet},
  journal = {Science},
  volume  = {323},
  pages   = {915--919},
  year    = {2009},
  doi     = {10.1126/science.1166767}
}

@article{Yu2010,
  author  = {Yu, X. Z. and Onose, Y. and Kanazawa, N. and Park, J. H. and Han, J. H. and Matsui, Y. and Nagaosa, N. and Tokura, Y.},
  title   = {Real-space observation of a two-dimensional skyrmion crystal},
  journal = {Nature},
  volume  = {465},
  pages   = {901--904},
  year    = {2010},
  doi     = {10.1038/nature09124}
}

@article{NagaosaTokura2013,
  author  = {Nagaosa, N. and Tokura, Y.},
  title   = {Topological properties and dynamics of magnetic skyrmions},
  journal = {Nat. Nanotechnol.},
  volume  = {8},
  pages   = {899--911},
  year    = {2013},
  doi     = {10.1038/nnano.2013.243}
}

@article{Jiang2017,
  author  = {Jiang, W. and Zhang, X. and Yu, G. and Zhang, W. and Wang, X. and Jungfleisch, M. B. and Pearson, J. E. and Cheng, X. and Heinonen, O. and Wang, K. L. and Zhou, Y. and Hoffmann, A. and te Velthuis, S. G. E.},
  title   = {Direct observation of the skyrmion {H}all effect},
  journal = {Nat. Phys.},
  volume  = {13},
  pages   = {162--169},
  year    = {2017},
  doi     = {10.1038/nphys3883}
}

\newpage
\foreach \x in {1,...,3}
{%
\clearpage
\includepdf[pages={\x}]{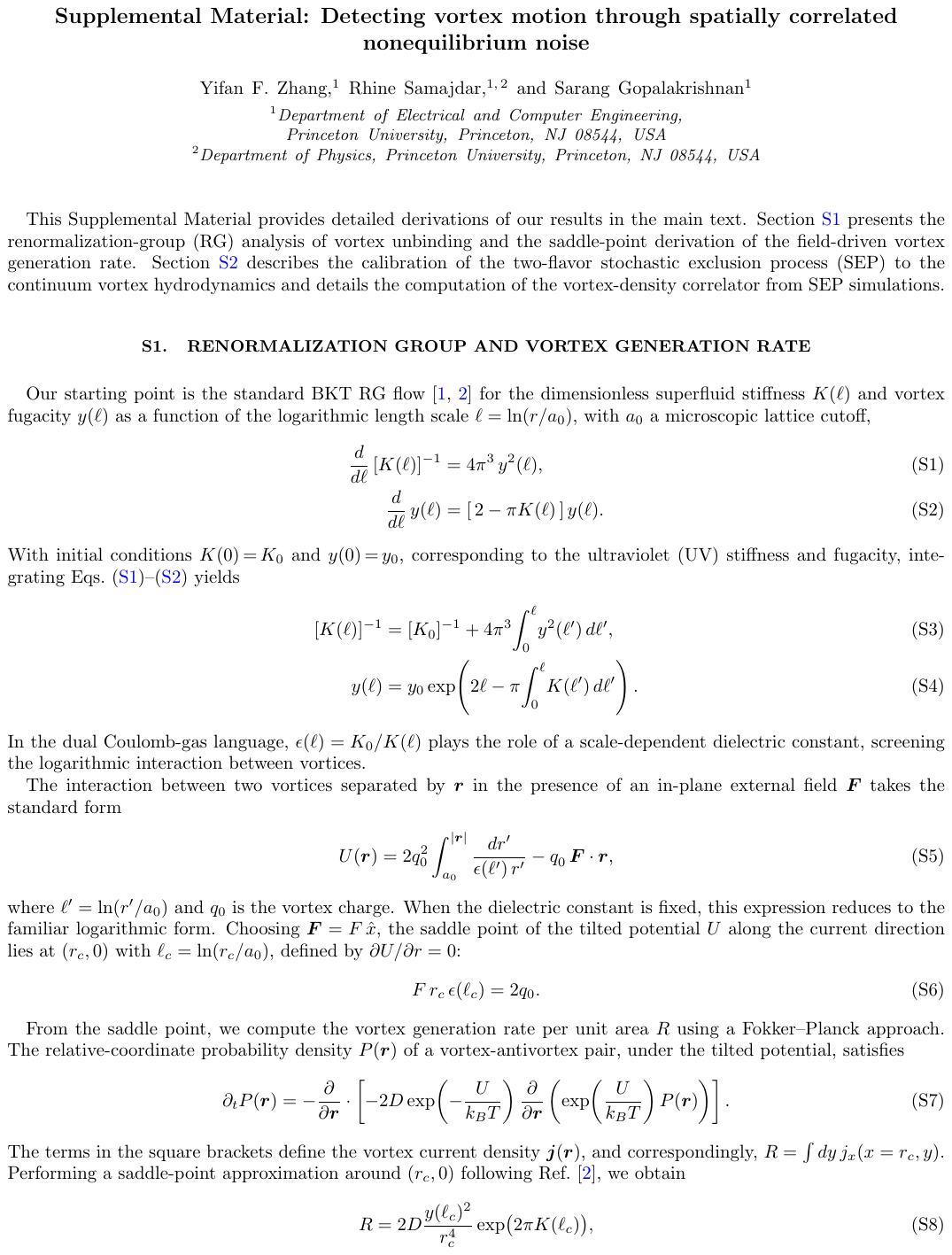} 
}

\end{document}